%% file: main.tex
\documentclass[aps,nopacs,twocolumn,preprintnumbers,notitlepage,amsmath,amssymb,superscriptaddress]{revtex4-2}
\usepackage[utf8]{inputenc}
\usepackage{amsmath}
\usepackage{amssymb}
\usepackage{graphicx}
\usepackage{import}
\usepackage[dvipsnames]{xcolor}
\usepackage{tikz}
\usepackage{makecell}
\usepackage{array}
\usepackage{booktabs}
\usepackage{bm}
\usepackage{placeins}
\usepackage{multirow}
\usepackage{hyperref}
\usepackage{adjustbox}
\usepackage[capitalise]{cleveref} 

\graphicspath{{Images/}{figs/}}

\newcommand{\mup}{\mu^\mathrm{(p)}}
\newcommand{\mus}{\mu^\mathrm{(s)}}
\newcommand{\mupt}{\tilde{\mu}^\mathrm{(p)}}
\newcommand{\must}{\tilde{\mu}^\mathrm{(s)}}
\newcommand{\rhoz}[1]{\rho_{0 #1}}
\newcommand{\ofrt}{\left( \mathbf{r}, t \right)}

\newcommand{\redint}{\tilde{\Lambda}}
\newcommand{\su}[1]{{\mathrm{s}(#1)}}
\newcommand{\pr}[1]{{\mathrm{p}(#1)}}
\newcommand{\Dp}{D_\mathrm{p}}
\newcommand{\tablegraphics}[2]{ \adjustbox{valign=c}{\includegraphics[height=#1]{#2}} }
\newcommand{\musovmup}[1]{ \frac{\mus_{#1}}{\mup_{#1}} }
\newcommand{\mupovmus}[1]{ \frac{\mup_{#1}}{\mus_{#1}} }

\begin{document}
\title{Interaction-motif-based classification of self-organizing metabolic cycles}
\author{Vincent Ouazan-Reboul}
\affiliation{Max Planck-Institute for Dynamics and Self-Organization, Am Fassberg 17, D-37077, G\"{o}ttingen, Germany}
\author{Ramin Golestanian}
\affiliation{Max Planck-Institute for Dynamics and Self-Organization, Am Fassberg 17, D-37077, G\"{o}ttingen, Germany}
\affiliation{Rudolf Peierls Centre for Theoretical Physics, University of Oxford, OX1 3PU, Oxford, UK}
\author{Jaime Agudo-Canalejo}
\affiliation{Max Planck-Institute for Dynamics and Self-Organization, Am Fassberg 17, D-37077, G\"{o}ttingen, Germany}

\begin{abstract}
Particles that are catalytically-active and chemotactic can interact through the concentration fields upon which they act, which in turn may lead to wide-scale spatial self-organization. When these active particles interact through several fields, these interactions gain an additional structure, which can result in new forms of collective behaviour. Here, we study a mixture of active species which catalyze the conversion of a substrate chemical into a product chemical, and chemotax in concentration gradients of both substrate and product. Such species develop non-reciprocal, specific interactions that we coarse-grain into attractive and repulsive, which can lead to a potentially complex interaction network. We consider the particular case of a metabolic cycle of three species, each of which interacts with itself and both other species in the cycle. We find that the stability of a cycle of species that only chemotax in gradients of their substrate is piloted by a set of two parameter-free conditions, which we use to classify the low number of corresponding interaction networks. In the more general case of substrate- and product-chemotactic species, we can derive a set of two high-dimensional stability conditions, which can be used to classify the stability of
all the possible interaction networks based on the self- and pair-interaction motifs they contain. The classification scheme that we introduce can help guide future studies on the dynamics of complex interaction networks and explorations of the corresponding large parameter spaces in such metabolically active complex systems.

\end{abstract}

\maketitle

\section{Introduction}

Chemotactic particles, which develop force-free motion in response to gradients, have been shown to develop effective interactions through the fields upon which they act, independently of the particular mechanism through which they move \cite{Golestanian2019phoretic}. Such interactions have for instance been observed in diffusiophoretic \cite{varma2018Clusteringinduced,Stark2018,saha2014clusters} and thermophoretic \cite{schmidt2019Lightcontrolled,golestanian2012Collective,cohen2014Emergent} colloids, as well as for chemotactic microorganisms \cite{keller1970Initiation,gelimson2015Collective}.
Field-mediated active interactions have the particularity of being nonreciprocal, meaning that the response of a particle A to the presence of another particle B is different from the response of B to A \cite{soto2014SelfAssembly,soto2015Selfassembly,sengupta2011Chemotactic,meredith2020predator}. This feature can lead to new forms of collective behaviour as compared to reciprocally-interacting systems \cite{saha2020Scalar,fruchart2021Nonreciprocal}. For example, binary mixtures of catalytically active particles may spontaneously form self-propelled clusters \cite{agudo-canalejo2019Active}. The models describing active phoretic particles can be extended to the case in which particles act upon and interact through several chemical fields, which introduces a nontrivial topology to the network of interactions among different particle species. This interaction topology can allow the self-organization of a mixture of self-repelling species \cite{ouazanreboul2023network}, which is not possible for simpler interaction schemes, and can lead to super-exponential aggregation of complementary catalysts, which could have been relevant in the emergence of living matter \cite{ouazanreboul2023Self-organization}.

Taking into account the presence of several chemical fields leads to an explosion in the number of possible interaction patterns that active phoretic species can develop, which in turn makes the determination of the conditions in which mixtures of such species can self-organize rather challenging. To overcome this issue, inspiration can be found from the methods used for tackling the protein folding problem, which consists in understanding and characterizing the process through which an initially linear chain of amino acids folds into a dense three-dimensional structure \cite{dill2008Protein}, and determining which structure a given sequence folds into.
One of the challenges that is encountered in this setting is that proteins are composed of 20 possible amino acid species, leading to a number of possible sequences going as $ 20^n $ for a chain of $ n $ amino acids, thus making the enumeration of biologically-relevant chains of hundreds of amino acids combinatorially complex. In order to decrease this number, one approach is to coarse-grain the amino acids into a smaller number of categories that have simple interaction rules. This is the basis of the celebrated HP lattice model \cite{dill1985Theory,phillips2012Physical}, which considers two
categories of hydrophobic and polar amino acids, although it has also been extended to larger alphabets \cite{hoque2009Extended,chan1999Folding,riddle1997Functional}. While current state of the art methods can leverage abundant computational power and large data sets, which allows for accurate predictions of the structure of biological proteins \cite{jumper2021Highly}, simple lattice models have impressive predictive power \cite{Li1996,Li1998}, with the hydrophobic or polar nature of the residues having been shown to have a strong influence on protein structure independently of their particular nature \cite{lim1989Alternative}. Moreover, simple proteins having been successfully designed based on the binary alphabet used in the HP model \cite{kamtekar1993Protein}.

In this work, we use a similar approach to determine the ability of catalytic species involved in small metabolic cycles to self-organize, coarse-graining their interactions into attractive ``A'' or repulsive ``R'' but with the important new element of non-reciprocity. We are able to systematically classify the (linear) stability of homogeneous mixtures of catalytic particles participating in metabolic cycles, and thus their tendency to spatially self-organize, based on which interaction motifs they contain. The article is structured as follows. Section \ref{sec:summary} describes the basic framework and summarizes the classification of the metabolic networks. Section \ref{sec:model} describes the model in detail, and our method for analyzing the stability of
metabolic cycles. In Section \ref{sec:phase_diag_3_1mob}, we classify the stability of cycles composed of species with a single chemotactic coefficient corresponding to the substrate of the reaction they catalyze, which can be done using a simple, parameter-free instability criterion. Finally, in Section \ref{sec:gen_class} we generalize this classification to species that are chemotactic to both their substrate and their product, by deriving two instability criteria and determining which interaction motifs need to be present for these instability conditions to be satisfied. Section \ref{sec:discuss} contains some discussions while some of the details of the calculations are relegated to the Appendices.

\begin{figure}[t]
    \begin{center}
        \includegraphics[width=1.0\columnwidth]{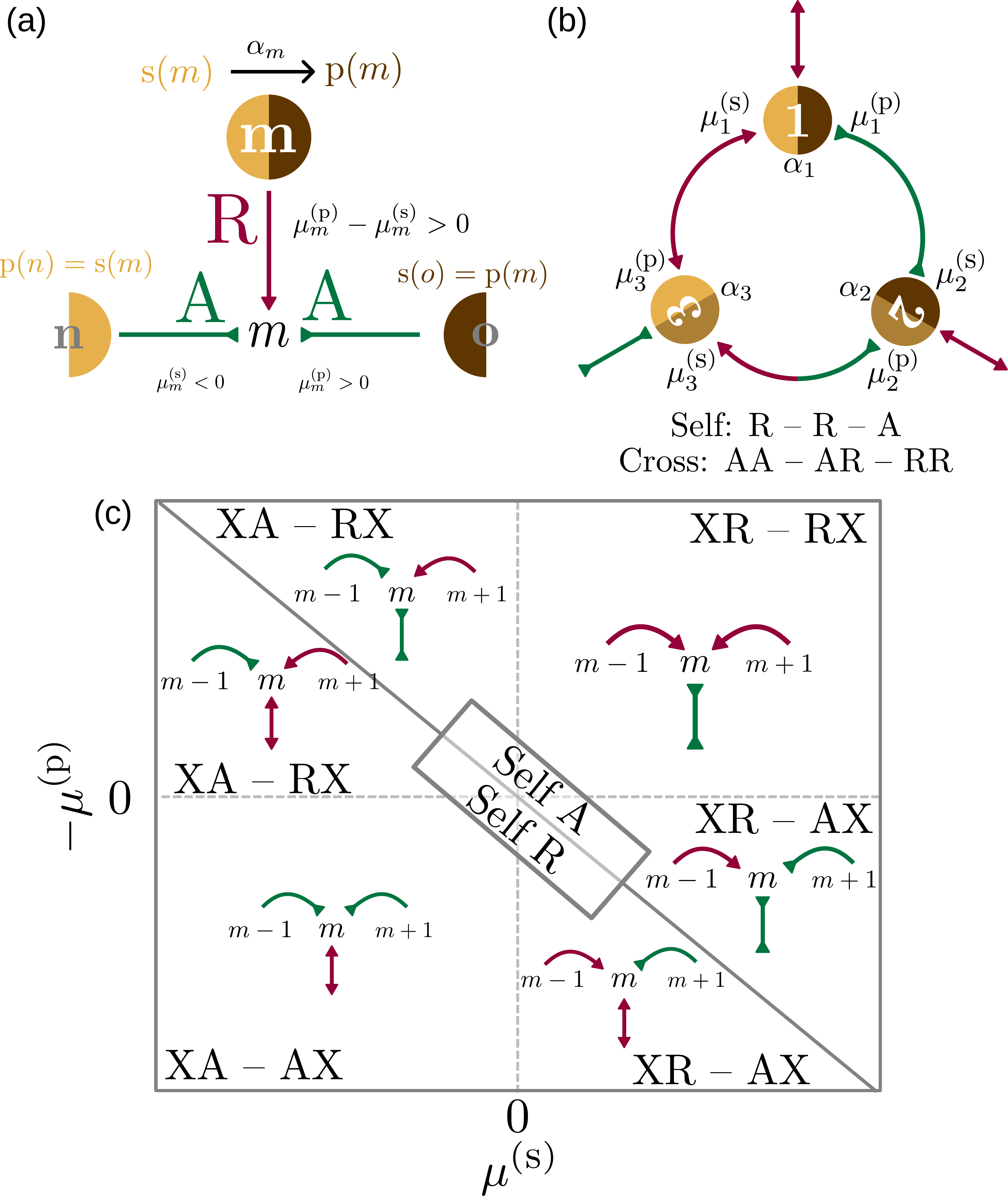}
    \end{center}
   \vskip-0.5cm
    \caption{
        (a) Specific interaction between catalytically active species. Particles of species $ m $ convert a substrate chemical into a product chemical at a rate $ \alpha_m $, and chemotax in concentration gradients of these two chemicals. These particles develop interactions with other catalytic species which modify the concentrations of their substrate (e.g.~species $ n $) and product (e.g.~species $ o $). 
        (b) Metabolic cycle of three species, which interact with themselves and their neighbour in the cycle. Note the nonreciprocity of the 2-3 pair interaction.
        (c) All possible interaction motifs between a given species and its neighbors in the cycle, determined by its substrate mobility $ \mus $ and its product mobility $ \mup $.
        ``A'' denotes self-attracting species and ``R'' denotes self-repelling species. ``X'' can correspond to any sense of interaction.
    }
    \label{fig:act_mob_cycle}
\end{figure}

\section{Model overview and summary of results}
\label{sec:summary}

We study self-organization of catalytically active particles belonging to $ M $ distinct species, and $K$ chemical fields upon which these particles act. In particular, each particle of species $m$ converts a substrate chemical, with index $ k=\su{m} $, into a product chemical  $ k'=\pr{m} $ at a rate $ \alpha_m
$  (\cref{fig:act_mob_cycle}(a)). The functions $\su{m}$ and $\pr{m}$ therefore map each catalytic species index to the index of the substrate and product chemical, respectively, and define the topology of the \emph{catalytic network}.
Assuming that catalysts of species $ m $ have a concentration $ \rho_m $, we can then write the evolution equations for
the concentrations of  the $ K $ chemical fields they produce and consume as:
\begin{equation}
    \label{eq:c_time_ev}
    \partial_t c_k (\mathbf{r}, t) 
    = 
    D^{(k)} \bm{\nabla}^2 c_k
    + \sum_{m} \left( \delta_{k, \pr{m}} - \delta_{k, \su{m}} \right)  \alpha_m \rho_m,
\end{equation}
where $ D^{(k)} $ is the diffusion coefficient of chemical species $ k $, and  $ \delta_{m,n} $ is the Kronecker symbol,
verifying $ \delta_{m,n} = 1 $ if $ m = n $ and 0 else.

The active particles we consider are also taken to be chemotactic for both their substrate and their product.
In a concentration gradient of its substrate $ \nabla c_\su{m} $, a particle of species $ m $ develops a velocity going
as $ - \mus_m \nabla c_\su{m} $, which takes it towards the high (respectively low) concentrations of its substrate if $
\mus_m $ is negative (respectively positive).
The same is true for gradients of the product of species $ m $, to which a product mobility $ \mup_m $, also of
arbitrary sign, and independent from $ \mus_m $, is associated.
We can then write the following continuity equation for the concentrations of the active species $ \rho_m $:
\begin{equation}
    \label{eq:rho_time_ev}
    \partial_t \rho_m (\mathbf{r}, t) 
    =
    \bm{\nabla} \cdot
    \left[
        D_\mathrm{p} \bm{\nabla} \rho_m
        +
        \left( \mus_m \bm{\nabla} c_\su{m} +
            \mup_m \bm{\nabla} c_\pr{m}
        \right)
        \rho_m 
    \right],
\end{equation}
where we take all catalysts to have the same diffusion coefficient $ \Dp $.

From \cref{eq:c_time_ev,eq:rho_time_ev}, it can be deduced that the catalytic species develop interactions mediated by
the concentration fields of their respective substrates and products: through their chemotactic mobilities, the active
particles we consider are able to  develop induced velocities in response to chemical gradients created by the catalytic
activity of other particles.
Effective interactions relying on this mechanism have been previously shown to lead to self-organization of catalytic mixtures
\cite{agudo-canalejo2019Active,ouazan-reboul2021Nonequilibrium,ouazanreboul2023Self-organization,ouazanreboul2023network}.
Here, because we study species which act upon and react to a restricted subset of concentration fields, the emergent
interactions are specific: a given species $ m $ only responds to concentration gradients of its substrates and
products, and thus only species which act on these particular concentration fields will elicit a induced response in $ m
$ (\cref{fig:act_mob_cycle}(a)).
Moreover, these interactions are nonreciprocal: in general, the velocity response of species $ m $ to species $ n
$ is different from the response of $ n $ to $ m $ (\cref{fig:act_mob_cycle}(b)).
Based on the choice of its mobilities, each catalyst species then develops a specific pattern of induced velocities to
the other catalytic species, which determines the \emph{interaction network} among species.

\input{Tables/cycle_stability.tex}

In this work, we seek to understand how different combinations of interaction patterns can lead to self-organization of
catalytic mixtures whose components interact according to a network of specific, nonreciprocal interactions.
We focus our study on the particular case of three species which are arranged into a model metabolic cycle
(\cref{fig:act_mob_cycle}(b)), in which the substrate of species 1 is the product of species 3 and its product is the
substrate of species 2, whose product is in turn the substrate of species 3.
Catalytic species involved in such a cycle develop interactions with both other species and themselves, according to a
set of six patterns shown in \cref{fig:act_mob_cycle}(c).
In the later sections, we will systematically study the stability of all interaction pattern combinations, and find that
any metabolic cycle of three species can be mapped onto nine ``elementary'' sets of networks, each of which belongs to 
one of five stability classes. The stability of all elementary network sets is listed in \cref{tab:elem_nets}, with the possible stability classes comprising the following cases:
\begin{itemize}
    \item Always unstable: cycles of three self-attracting species which self-organize no matter the choice of parameters.
    \item Type-I unstable: cycles which involve self-attracting species, and which can self-organize if self-attraction overcomes self-repulsion.
    \item Type-IIa (or strongly) unstable: cycles of self-repelling species involving at least one instability-favouring pair.
    \item Type-IIb (or weakly) unstable: cycles of self-repelling species which include at least one pair which can be instability-favouring if some constraint on the mobilities of the active species is satisfied.
    \item Always stable: cycles of self-repelling species which do not include instability-favoring pair motifs, and cannot self-organize.
\end{itemize}

\section{Model description}
\label{sec:model}

We consider a set of $ M = 3 $ catalytic species involved in a metabolic cycle as described in the previous section. Due to this cycle structure, there are $ K=3 $ chemical fields through which the catalysts interact, and we choose the convention that $\su{m}=\pr{m-1}=m$. Note that, throughout this work, we will use periodic indices, so that species 0 is species 3 and species 4 corresponds to species 1, and so on.

In this setting, we perform a linear stability analysis on \cref{eq:c_time_ev,eq:rho_time_ev} (see Appendix \ref{sec:app_lin_stab}) and find that the stability of the cycle is set by the eigenvalue equation
\begin{equation}
    \label{eq:stability}
    \lambda \delta \rho_m = - \sum_{n=1}^3 \Lambda_{m,n} \delta \rho_n,
\end{equation}
with $\delta \rho_n$ being the perturbation of the concentration of species $n$.
The catalytic species involved in the metabolic cycle then undergo spatial self-organization through a system-wide instability if $ {\rm Re}(\lambda) > 0 $, the conditions for which we seek to uncover in the rest of this work. \cref{eq:stability} involves $ \bm{\Lambda} $, which is the matrix of effective interactions between the active species, and has coefficients
\begin{equation}
    \label{eq:ints}
    \Lambda_{m,n} = 
    \begin{cases}
        \alpha_{m-1} \must_m \rho_{0m}  \text{ if } n = m-1, \\
        \alpha_{m} \left(\mupt_m - \must_m\right) \rho_{0m}  \text{ if } n = m, \\
        - \alpha_{m+1} \mupt_m \rho_{0m}m  \text{ if } n = m+1.
    \end{cases}
\end{equation}
Here, we have set $ \must_m = \mus_m / D^{(m)} $ and $ \mupt_m = \mup_m / D^{(m+1)} $, and $\rho_{0m}$ is the density of species $m$ in the homogeneous state. The response of species $ m $ to species $ n $ is attractive if $ \Lambda_{m,n} < 0 $, and repulsive if $ \Lambda_{m,n}
> 0$.
We can then characterize the structure of the interactions of a metabolic cycle via the signs of the
self- and pair-interactions of its constituent species, the encoding of which is shown in \cref{fig:act_mob_cycle}(b): a repulsive
interaction is denoted by the letter ``R'', and an attractive interaction by the letter ``A''.
In the example of \cref{fig:act_mob_cycle}(b), the self-interaction pattern is written as R--R--A, meaning that species
1 and 2 are self-attracting, and species 3 is self-repelling. The pair interactions, meanwhile, read AA--AR--RR, which denotes the fact that species 1 and 2 attract each other, species 2 chases species 3, and species 1 and 3 both repel each other.

According to \cref{eq:ints}, the self-interaction of a given species $ m $ can be expressed as a linear combination of
its cross-interactions with other species.
This constrains the signs of the triplet $ \left( \Lambda_{m-1,m}, \Lambda_{m,m}, \Lambda_{m,m+1} \right) $ to belong to
one of the six, rather than the theoretically possible $ 2^3 = 8 $, patterns shown in \cref{fig:act_mob_cycle}(c).
Any possible interaction network for a metabolic cycle of three species can then be built by independently choosing one
of these six patterns for each species.
In the rest of this work, we classify the stability behaviour of all such possible $ 6^3 = 216 $ networks.

\section{Phase diagram for product-insensitive species}
\label{sec:phase_diag_3_1mob}

The number of possible interaction combinations can be reduced by setting the product mobilities of all species to zero, i.e. $ \mup_m = 0$, which sets $ \Lambda_{m,m+1} = 0 $ for all species.
In this case, catalysts of a given species $ m $ can only interact with catalysts of the same species or the previous
species $ m-1 $.
Additionally, the self-interaction now satisfies $ \Lambda_{m,m} = - \frac{\alpha_m}{\alpha_{m+1}} \Lambda_{m,m-1} $,
meaning that the self-interaction of a given species is of the opposite sign as its response to the previous species.
Only two interaction motifs are then possible in this case, one being self-attracting and repelled by the previous species,
and the other, self-repelling and attracted by the previous species. This makes for a total number of possible interaction networks of $ 2^3 = 8$, which are easily enumerated.

We can determine the stability of each network of purely substrate-sensitive interactions by carrying out the linear
stability analysis, which in this particular case yields  parameter-independent instability conditions, and determining how
these conditions intersect with regions of the parameter space corresponding to the different interaction patterns.
The eigenvalues for strictly substrate-sensitive species are found as
\begin{widetext}
\begin{equation}
    \label{eq:omob_eig_reduced}
    \lambda_\pm =\left| \Lambda_{1,1} \right|\left[
    - \frac{1}{2} (\redint_1 + \redint_2 + \redint_3)
    \pm \frac{1}{2}
    \sqrt{
        \left( \redint_1 + \redint_2 + \redint_3 \right)^2 -
        4  \left( \redint_1 \redint_2 + \redint_2 \redint_3 + \redint_1 \redint_3 \right)
    }\right].
\end{equation}
\end{widetext}
Here, we have normalized the self-interactions with  $ \Lambda_{1,1} $,  such that $\redint_{m} = 
\frac{\Lambda_{m,m}}{\left| \Lambda_{1,1} \right|}$ is the ratio of the self-interaction of species $m$ to the magnitude of the
self-interaction of species 1.
From \cref{eq:omob_eig_reduced}, two instability conditions can be deduced, which we respectively name type-I and type-II:
\begin{equation}
    \label{eq:omob_first_order_cond}
  \text{type-I:    } \;\; \redint_1 + \redint_2 +\redint_3 < 0,
\end{equation}
and
\begin{equation}
    \label{eq:omob_second_order_cond}
  \text{type-II:    }  \begin{cases}
        \redint_1 + \redint_2 +\redint_3 > 0, \\
        \redint_1 \redint_2 + \redint_2 \redint_3 + \redint_1 \redint_3 < 0.
    \end{cases}
\end{equation}
The type-I condition \cref{eq:omob_first_order_cond} only contains self-interaction terms, and corresponds to having a
mixture that is self-attracting on average, as found in Refs.~\cite{agudo-canalejo2019Active} and
\cite{ouazan-reboul2021Nonequilibrium} for systems with a non-cyclic interaction topology.
In contrast, the type-II condition \cref{eq:omob_second_order_cond} includes terms involving pairs of particle species.
This condition allows for the mixture to be self-repelling on average, and instead requires pairs with opposite
self-interaction signs to have stronger self-interaction than pairs with equal self-interaction signs.

\begin{figure*}[t]
    \centering
    \includegraphics[width=\textwidth]{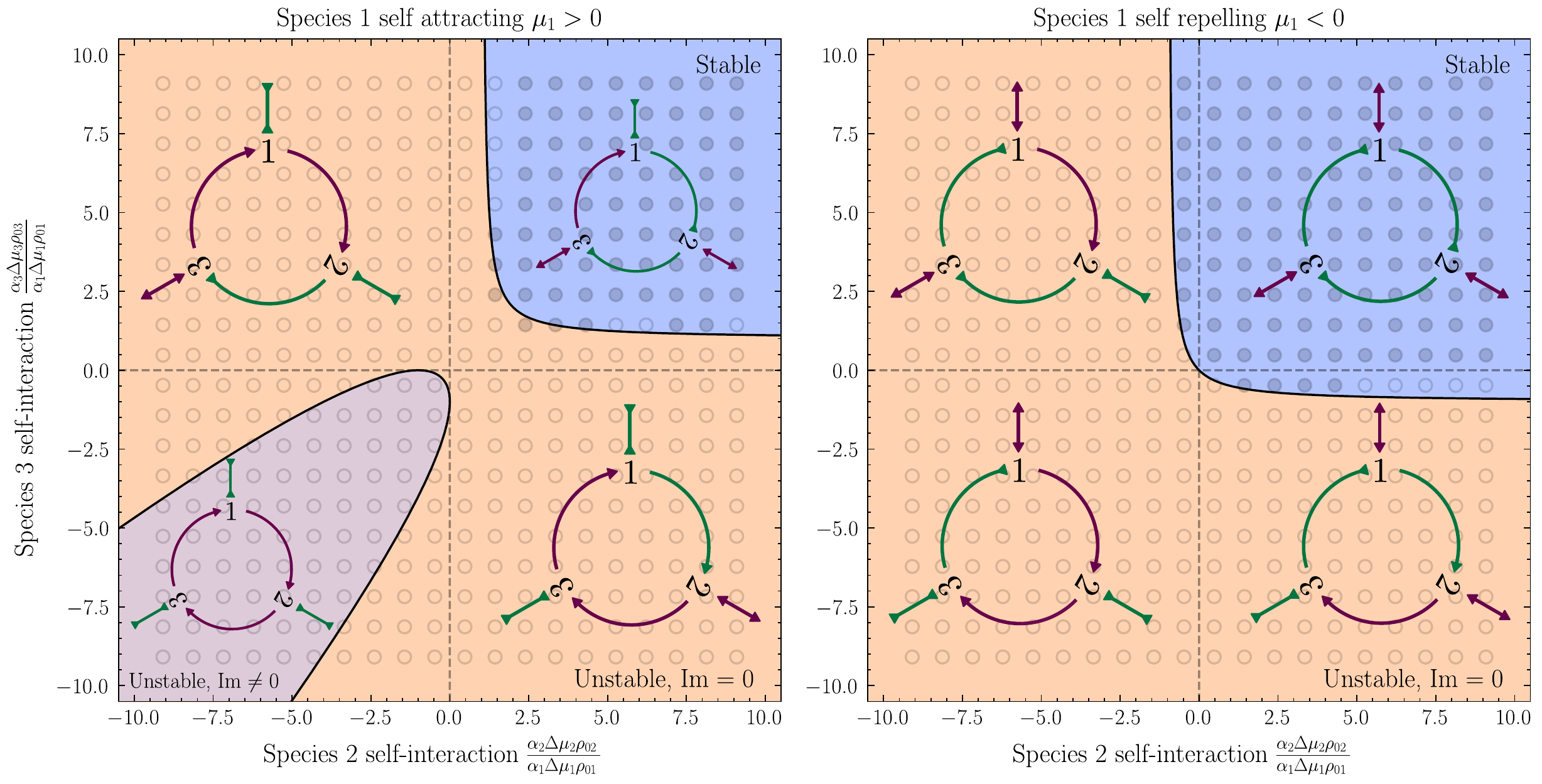}
    \caption{
        {Stability diagram for a cycle of three species chemotactic only to their 
        substrate.}
        Species 1 is self-attracting ($\mu_1>0$) in the left panel, and self-repelling ($\mu_1<0$) in the right panel.
        Complex eigenvalues  occur in the purple region. Grey circles correspond to the results of Brownian dynamics
        simulations, with empty and filled circles corresponding to stable and unstable homogeneous states,
        respectively. For the simulation parameters and the expressions defining the stability lines, see Appendix \ref{sec:app_simul}.
    }
    \label{fig:omob_ph_diag}
\end{figure*}

We can then study separately the cases where species 1 is self-attracting ($\redint_1=-1$) and self-repelling ($\redint_1=1$).
In the case of a self-attracting species 1, $ \redint_1 = -1 $, we rewrite
\cref{eq:omob_first_order_cond,eq:omob_second_order_cond} as inequalities of $ \redint_3 $ as a function of $ \redint_2 $.
It can be shown that, if $\redint_2 < 1$, the first and second order conditions are complementary, so that at least one is
always satisfied, and that the overall instability condition then writes:
\begin{equation}
    \label{eq:omob_sa_instab}
    \left\{
        \begin{aligned}
            &\redint_2 \geq  1,\\
            &\redint_3 < \frac{\redint_2}{\redint_2-1},
        \end{aligned}
    \right.
        \text{ or } \
        \redint_2 < 1.
\end{equation}
Furthermore, we can also write an oscillation condition by determining which parameters make \cref{eq:omob_eig_reduced}
complex with a positive real part.
Parameters which make the term under the square root in \cref{eq:omob_eig_reduced} also satisfy the type-I
instability condition, leading to the oscillation condition
\begin{equation}
    \label{eq:omob_oscs}
    \begin{cases}
        &\redint_2 < 0, \\ 
        &\redint_3 \in 
        \left[
            -(1+\sqrt{-\redint_2})^2, -(1-\sqrt{-\redint_2})^2
        \right].
    \end{cases}
\end{equation}
We plot the corresponding stability phase diagram in $ (\redint_2, \redint_3) $ coordinates in the left panel of
\cref{fig:omob_ph_diag}, and find that cycles of purely substrate-sensitive species are always unstable if they
contain two or three self-attracting species, while they can be either stable or unstable if they contain one
self-attracting and one self-repelling species.
The oscillation condition \cref{eq:omob_oscs}, meanwhile, is only compatible with cycles of three self-attracting
species (\cref{fig:omob_ph_diag}, purple region).
These predicted stability diagram was confirmed by particle-based Brownian dynamics simulations (see Appendix \ref{sec:app_simul}). In \cref{fig:omob_ph_diag}, filled circles correspond to simulations that remained homogeneous, while empty circles correspond to simulations that displayed clustering. Oscillations were never observed in simulations, which we ascribe to the fact that
the real part of the complex eigenvalue is always larger than its imaginary part, meaning that the system is out of the linear-perturbation regime by the time a full oscillation period is completed.

We now consider species 1 to be self-repelling, $ \redint_1 = 1 $. Using similar calculations as in the self-attracting case, we find the instability condition
\begin{equation}
    \label{eq:omob_sr_instab}
    \left\{
        \begin{aligned}
            &\redint_2 \geq -1,\\
            &\redint_3 < -\frac{\redint_2}{\redint_2+1},
        \end{aligned}
    \right.
        \text{or} \
        \redint_2 < -1.
\end{equation}
Plotting the corresponding phase diagram in the right panel of \cref{fig:omob_ph_diag}, we once again find that cycles
of two self-attracting, product-insensitive species are always unstable and that cycles including only one
self-attracting species can either be stable or unstable.
Finally, we find that a cycle composed of three self-repelling species verifying $ \mup = 0 $ is always stable.
The oscillation condition obtained with $ \redint_1 = 1 $ is incompatible with the set of inequalities
\cref{eq:omob_sr_instab}, confirming that all three species need to be self-attracting  for the eigenvalue to be
complex. These predictions were again successfully verified by the results of our Brownian dynamics simulations.

\section{Classification of generic cycles}\label{sec:gen_class}

\subsection{Type-I instabilities in generic cycles}\label{sec:typeIgen}

We now consider the general case with both nonzero substrate and product mobilities.
In this more general case, networks can be built by choosing three of any of the six
 interaction patterns from \cref{fig:act_mob_cycle}(c), and there are then 
$6^3 = 216$ possible interaction networks. 
We show here that the presence or absence of certain motifs in the interaction network 
can be used to infer the stability of the full system by using a semi-diagrammatic approach.
Similarly to cycles of product-insensitive species, we first calculate the non-null eigenvalues, which are found as 
\begin{widetext}
\begin{equation}
    \label{eq:3_eigval}
    \lambda_\pm 
    =
    - \frac{1}{2}
    \sum\limits_{m=1}^3 
    \Lambda_{m,m}
    \pm
    \frac{1}{2}
    \sqrt{
        \left( 
            \sum\limits_{m=1}^3 
            \Lambda_{m,m}
        \right)^2
        -  
        4 \sum\limits_{m=1}^3 
        \left( 
            \Lambda_{m,m} \Lambda_{m+1,m+1}
            - \Lambda_{m+1,m} \Lambda_{m,m+1}.
        \right)
    }.
\end{equation}
\end{widetext}
The homogeneous state is linearly unstable when the real part of the largest eigenvalue in \cref{eq:3_eigval}
becomes positive, which leads to a type-I instability when
\begin{equation}
    \label{eq:gen_instab_1}
    \sum_{m=1}^{3} \Lambda_{m,m} < 0,
\end{equation}
effectively the same condition as \cref{eq:omob_first_order_cond}, and corresponds to an overall
self-attracting metabolic cycle.

Similarly to the case with substrate sensing only, any cycle of three self-attracting species is necessarily unstable
according to \cref{eq:gen_instab_1}.
Cycles including at least one self-attracting and one self-repelling species can be either stable or unstable according
to the type-I condition (\cref{tab:elem_nets}, first two lines).
The presence of at least one self-attracting species is then sufficient to conclude on the possible stability behavior
of a given cycle, independent of the rest of the interactions, which allows us to classify the 189 networks with at
least one self-attracting species.

\input{Tables/pair_stability.tex}

\subsection{Stability of self-repelling species pairs}\label{sec:pair_class}

The 27 cycles that do not contain any self-attracting species can only exhibit a type-II instability, which corresponds to the following condition:
\begin{equation}
    \label{eq:gen_instab_2}
    \sum\limits_{m=1}^{3} \left( \Lambda_{m,m} \Lambda_{m+1, m+1} - \Lambda_{m,m+1} \Lambda_{m+1, m}  \right) < 0.
\end{equation}
This inequality involves the sum of three contributions, each of which corresponds to one of the three species pairs.
We first determine which of the $ 3^2 = 9 $ possible pairs of self-repelling species provide negative contributions to
the sum written in \cref{eq:gen_instab_2} (see Appendix \ref{sec:app_pair_stab}).
To do so, we expand the terms of the sum as
\begin{equation}
    \label{eq:instab_2_terms}
    \sum\limits_{m=1}^{3} \alpha_m \alpha_{m+1} \Delta_{m, m+1} \rhoz{m} \rhoz{m+1} < 0,
\end{equation}
using the following definition
\begin{equation}
    \label{eq:type_2_terms}
    \Delta_{m,m+1} = \must_m\must_{m+1} + \mupt_m \mupt_{m+1} - \must_m\mupt_{m+1},
\end{equation}
and find the conditions on the mobilities of each of the self-repelling species which leads to $ \Delta_{m,m+1} < 0 $ (see
Appendix \ref{sec:app_pair_stab} for the details of the derivation).

We compile the results in \cref{tab:motif_stab}. Out of the nine pairs of
self-repelling species, five are always stabilizing, one is always destabilizing, and three are conditionally destabilizing, meaning
that they favour instabilities if a certain inequality on the chemotactic mobilities of their constituent species is satisfied.
Pairs that develop chasing interactions are all stability-favouring, which can be predicted from
\cref{eq:gen_instab_2} with all self-interactions $ \Lambda_{m,m} $ positive.
On the other hand, species pairs which favor instability all interact reciprocally and involve some degree of repulsion,
which tends to make the pair more destabilizing when its magnitude is increased.
This is exemplified by the fact that the only pair which is always instability-favouring, whose interactions can be described as
XR--AA--RX, involves two species which reciprocally attract and are repelled by their other neighbour.

\subsection{Stability of cycles of self-repelling species}\label{sec:gen_typeII}

Using the information gathered on the pairs of self-repelling
species, we systematically classify the remaining 27 networks of only self-repelling species.
We further reduce that number to seven ``elementary'' networks onto which any cycle of self-repelling species can be
mapped through symmetry operations, using the invariance of the eigenvalue \cref{eq:3_eigval} under cyclic swap of
the catalytic species, and the invariance of \cref{eq:gen_instab_2} under mirror symmetry around one pair (see Appendix \ref{sec:app_sym_ops}). We then classify the stability of each elementary network based on which species pair they contain.

We find that two networks contain the always destabilizing pair XR -- AA -- RX \cref{tab:motif_stab}, and are thus unstable
if the magnitude of the activities, mobilities, or homogeneous concentrations associated to it are tuned so that the
associated term in \cref{eq:gen_instab_2} overcomes the stabilizing term from the other pairs.
As these cycles can be made unstable simply by adjusting the parameters in order to give enough weight to one-pair
terms, we call them type-IIa, or strongly, unstable.
Another cycle, whose pair interactions go as AA--AA--RR, contains three conditionally destabilizing pairs.
However, the inequalities to be satisfied for all three pairs to be stable are incompatible, and at least one of the
three is necessarily instability-favouring.
This cycle can then also be classified as type-IIa-unstable, with the only difference with the two previous one being
that the instability-favoring pair can be different based on the choice of mobilities.

Two more elementary cycles can be made unstable, but with stricter criteria than those of type-IIa.
These cycles only contain species pairs which are either stabilizing, or conditionally destabilizing.
In order for these to be unstable, two criteria then need to be fulfilled: the species mobilities must be chosen so that
at least one conditionally destabilizing pair is unstable, and enough weight must be given in \cref{eq:gen_instab_2} to this
pair, so that the whole cycle is unstable, a condition which we name type-IIb (or weakly) unstable networks.
Finally, two elementary cycles are only composed of stabilizing pairs, and thus cannot be made unstable.

\section{Discussion}\label{sec:discuss}

In this work, we have studied the behavior of a metabolic cycle of three catalytic species which chemotax in response to gradients of their substrates and products, and interact nonreciprocally owing to these two properties.
We have demonstrated that the interaction network between the species can be built by independently choosing one
interaction pattern per species, and that the stability of the resulting network can be determined by independently
considering single-species and species-pair motifs.
For the particular case of cycles composed of species with no chemotaxis in response to product gradients (only in response to substrate gradients), we have calculated a parameter-free stability line and used it to classify all the possible interaction networks in such cycles. We found that, for this reduced model, at least one self-attracting interaction motif must be present in
order to observe self-organization. For cycles of species chemotactic in response to both substrate and product gradients, we derived two instability criteria from which the stability behavior of any choice of interaction motifs can be determined.
The first condition (type-I) concerns the single-species interaction motifs, and translates the fact that the presence of
a self-attracting species is sufficient for a cycle to potentially be unstable. In the case in which a cycle is composed strictly of self-repelling species, a second condition (type-II) becomes relevant, which involves interaction motifs between pairs of species.
The contribution to all possible pair interaction motifs to this condition can be classified as either stabilizing, destabilizing, or conditionally destabilizing. In turn, this classification of pairs can be used to classify the stability of self-repelling cycles according to the pair motifs that they contain, and distinguish type-II unstable cycles as type-IIa, or strongly, unstable if they contain at least a destabilizing species pair, and type-IIb, or weakly, unstable if they only contain conditionally destabilizing
(and stabilizing) pairs.

We have restricted our investigation here to size-three metabolic cycles,  which are structures of small size and complexity compared to many biologically-relevant metabolic pathways. An immediate extension would then be to use the methods we developed for small cycles and to  apply them to larger chemical reaction networks, or to non-cycle geometries.
The most straightforward generalization is the study of metabolic cycles of more than three species, which have already
been shown to exhibit complex and cycle-size-dependent behaviour \cite{ouazanreboul2023Self-organization}.
Intriguingly, our motif-based classification has shown that, in a size-three cycle, pair interactions can lead to self-organization with more relaxed conditions than non-cyclic topologies. In particular, self-organization may occur even for systems in which all species are self-repelling, as studied in more detail in \cite{ouazanreboul2023network}.
Larger cycle sizes likely result in the emergence of higher-order terms in the linear stability analysis (for instance, triplets of species in a size-four cycle), which could further relax the instability condition and lead to new types of instability.
In the context of larger, more complex biochemical reaction networks, one possible approach could be to identify and analyze small key motifs in the reaction pathway, akin to the network motifs which are studied in systems biology \cite{alon2007Network,milo2002Network} and other disciplines \cite{stone2019Network}.
Finally, recent works have demonstrated that the spatial self-organization of enzymes can optimize the overall reaction rate of a pathway \cite{buchner2013Clustering} or be used as a method for regulating the output of a branch in a reaction network \cite{hinzpeter2019regulation}. Such effects could be investigated in the context of our model metabolic cycle, for instance by particularizing one of the reaction products and studying the effect of self-organization on its production rate.
Conversely, the spatial arrangement of catalysts can be designed to increase reaction yield \cite{xie2017Tandem}, implying that
reaction networks could be also be designed with a similar goal in mind by choosing catalysts which self-organize into a 
reaction-flux-optimizing structure.

\acknowledgments
This work has received support from the Max Planck School Matter to Life and the MaxSynBio Consortium, which are jointly funded by the Federal Ministry of Education and Research (BMBF) of Germany, and the Max Planck Society.

\section*{Appendices}
\appendix
\section{Linear stability analysis}\label{sec:app_lin_stab}

The stability of a size-three metabolic cycle can be determined by performing a linear stability analysis of
\cref{eq:c_time_ev,eq:rho_time_ev} as follows.
We consider a homogeneous steady state of \cref{eq:c_time_ev,eq:rho_time_ev}, with time- and space-independent concentrations
$c_{0k}$ for chemical species $k$ and $\rho_{0m}$ for catalytic species $m$.
We then perturb this steady state with perturbations $\delta \rho_n \ofrt$ and $\delta c_k \ofrt$, 
and develop \cref{eq:c_time_ev,eq:rho_time_ev} to the first order in perturbation.
With the quasi-steady-state assumption that $\partial_t \delta c_k =0$, corresponding to the limit of fast-diffusing chemicals species,
and using the Fourier mode decomposition $\delta \rho_n = \delta \rho_n (\bm{k}) \mathrm{e}^{\lambda t} \mathrm{e}^{i \bm{k} \bm{r}} $
and $\delta c_k = \delta c_k (\bm{k}) \mathrm{e}^{\lambda t} \mathrm{e}^{i \bm{k} \bm{r}} $,
we obtain the following equation:
\begin{equation}
    (\lambda + \Dp k^2) \delta \rho_m (\bm{k}) = - \sum_{n=1}^3 \Lambda_{m,n} \delta \rho_n (\bm{k}).
\end{equation}
It is easily seen that, if ${\rm Re}(\lambda) > 0$, then the $k=0$ mode, which corresponds to a system-wide perturbation,
is the most unstable one.
We then focus our analysis on that mode, and to simplify notation use $\delta \rho_m \equiv \delta \rho_m (k=0)$, which leads to
\cref{eq:stability}.

\section{Brownian dynamics simulation}\label{sec:app_simul}

In order to test the stability criteria for strictly substrate-chemotactic species given in 
\cref{eq:omob_sa_instab,eq:omob_sr_instab}, we perform Brownian dynamics simulations of our system.
We consider spherical catalysts of diameter $\sigma$, which we also take as the size scale for our simulations,
and assume that they isotropically convert their substrate into their product on their surface at a rate $\alpha$.

We first derive the expression for the velocity which a catalytically-active particle develops as a response to
the presence of another particle.
Using the same quasi-steady state approximation used in the linear stability analysis, we can write that
the concentration field associated with species $k$ follows at any time the Laplace equation
\begin{equation}
    D^{(k)} \nabla^2 c_k = 0,
\end{equation}
with boundary conditions obtained by balancing the reactive and diffusive fluxes at the surface of the catalysts.
A particle with index $i$ induces a velocity response for another particle $j$ by creating perturbations of 
the latter's substrate and product concentrations, which we write as $\delta c^{\su{j}}$ and $\delta c^{\pr{j}}$
respectively.
The corresponding velocity of $j$ in the presence of $i$ is then $\bm{v}_{ji}=\bm{v}_{ji}^\mathrm{s}+\bm{v}_{ji}^\mathrm{p}$ with $\bm{v}_{ji}^\mathrm{s} = - \mus_j \nabla \delta c_i^\su{j}$ and 
$\bm{v}_{ji}^\mathrm{p} = - \mus_j \nabla \delta c_i^\pr{j}$, which finally results in:
\begin{equation}
    \label{metheq:vel_resp}
    \bm{v}_{ji} = \frac{\Lambda_{j,i}}{4 \pi} \frac{\bm{r}_{ji}}{r_{ji}^3}
\end{equation}
where $\bm{r}_{ji}=\bm{r}_j-\bm{r}_i$, and $\Lambda$ is defined as in \eqref{eq:ints} except without the density $\rho_{0m}$.

The location $\bm{r}_i$ of catalyst $i$ then evolves according to the Langevin equation
\begin{equation}
    \label{metheq:langevin}
    \frac{\rm d}{{\rm d} t}\bm{r}_i(t) = \sum_{j \neq i} \bm{v}_{ij}  + \sqrt{2 \Dp} \bm{\eta}_i
\end{equation}
where $\bm{\eta}_i$ is a centred Gaussian white noise with intensity one.
We simulate a mixture of three species, each of which has a population $N_m$, yielding a set of $N_1 + N_2 + N_3$ Langevin
equations which we integrate using the forward Euler method with time step $\mathrm{d}t$ for a duration $t_\text{tot}$.
The three chemical species are assumed to have the same diffusion coefficient, $D^{(1)} = D^{(2)} = D^{(3)} = D$.
The quantities $\alpha_0$ and $\mu_0$ are arbitrary activity and mobility scales, from which a time scale
$\tau = \alpha_0 \mu_0 / (4 \pi D \sigma^3)$ and a diffusion coefficient scale $D_0 =  (4 \pi \sigma) / \alpha_0 \mu_0$
can be built.
The particles are simulated in a cubic box of size chosen so that the particles occupy a volume fraction $\Phi = 0.005$, with the 
interactions implemented according to the minimum image convention.
Finally, we simulate short-range repulsion between the catalytic particles by performing an overlap correction after each
time step using the elastic collision method described in \cite{strating1999Brownian}.

To generate the phase diagram shown in \cref{fig:omob_ph_diag}, we simulate a set of $N_1 = N_2 = N_3 = 500$ particles with the following parameters:
activity $\alpha / \alpha_0 = 1$, time step $\mathrm{d}t / \tau = 0.001$, simulation duration $t_\text{tot}/\tau = 600$,
noise intensity $\Dp / D_0 = 0.01$.
The mobility of species 1 is fixed at $\mus_1 / \mu_0 = \pm 0.33$, and the mobilities of species 2 and 3 are taken on a
20 by 20 grid with range $\mus_{2,3}/\mu_0 \in [-3, 3]$.

\section{Determination of pair instability conditions}\label{sec:app_pair_stab}
We use the fact that the three possible
interaction motifs for self-repelling species are each associated to certain constraints on the species mobilities (see
\cref{fig:act_mob_cycle}(c) for the mobility signs corresponding to the interactions):
\begin{itemize}
    \item XA--RX: As $ \mus_m < 0, \mup_m < 0, \mup_m > \mus_m $, we can write:
        \begin{equation}
            0 < \mupovmus{m} < 1 \\
            \label{eq:MM_cond}
        \end{equation}
    \item XR--AX: As $ \mus_m > 0, \mup_m > 0, \mup_m > \mus_m $, we can write:
        \begin{equation}
                0 < \musovmup{m} < 1
            \label{eq:PP_cond}
        \end{equation}
    \item XA--AX: As $ \mus_m < 0, \mup_m > 0$, we can write:
        \begin{equation}
            \mupovmus{m} < 0 \\
            \label{eq:MP_cond}
        \end{equation}
\end{itemize}
We then study the stability of all possible pairs of these motifs.
To do so, we first write the pair stability factor as given by \cref{eq:gen_instab_2}.
We write the condition $ \Delta_{m, m+1} < 0 $, subtract the negative terms so that the inequality only involves
positive quantities, and rewrite the resulting inequality into a condition involving the ratios $ \musovmup{m ~
\text{or} ~ m+1} $ and $ \mupovmus{m ~ \text{or} ~ m+1} $.
From the conditions in \cref{eq:MM_cond,eq:PP_cond,eq:MP_cond}, it can then be determined whether the motif is unstable.
For instance, for the interaction motif XA--RA--RX, the condition can be rearranged to have positive
left- and right-hand-side terms as $\mup_m \mup_{m+1} + \mus_m \mus_{m+1} < \mus_m \mup_{m+1} $.
This inequality can be further rearranged in terms of mobility ratios  under the previously enumerated constraints as $
\musovmup{m+1} + \mupovmus{m} < 1$.
Because $\musovmup{m+1} > 1$ and $ \mupovmus{m} > 0 $, this condition is impossible to realize: the motif is always
stable.
By systematically applying this recipe, we analyze the stability of all motifs, which is given in \cref{tab:motif_stab}.

\section{Mirror symmetry operation}\label{sec:app_sym_ops}
\begin{figure}[b]
\centering
\includegraphics[width=\columnwidth]{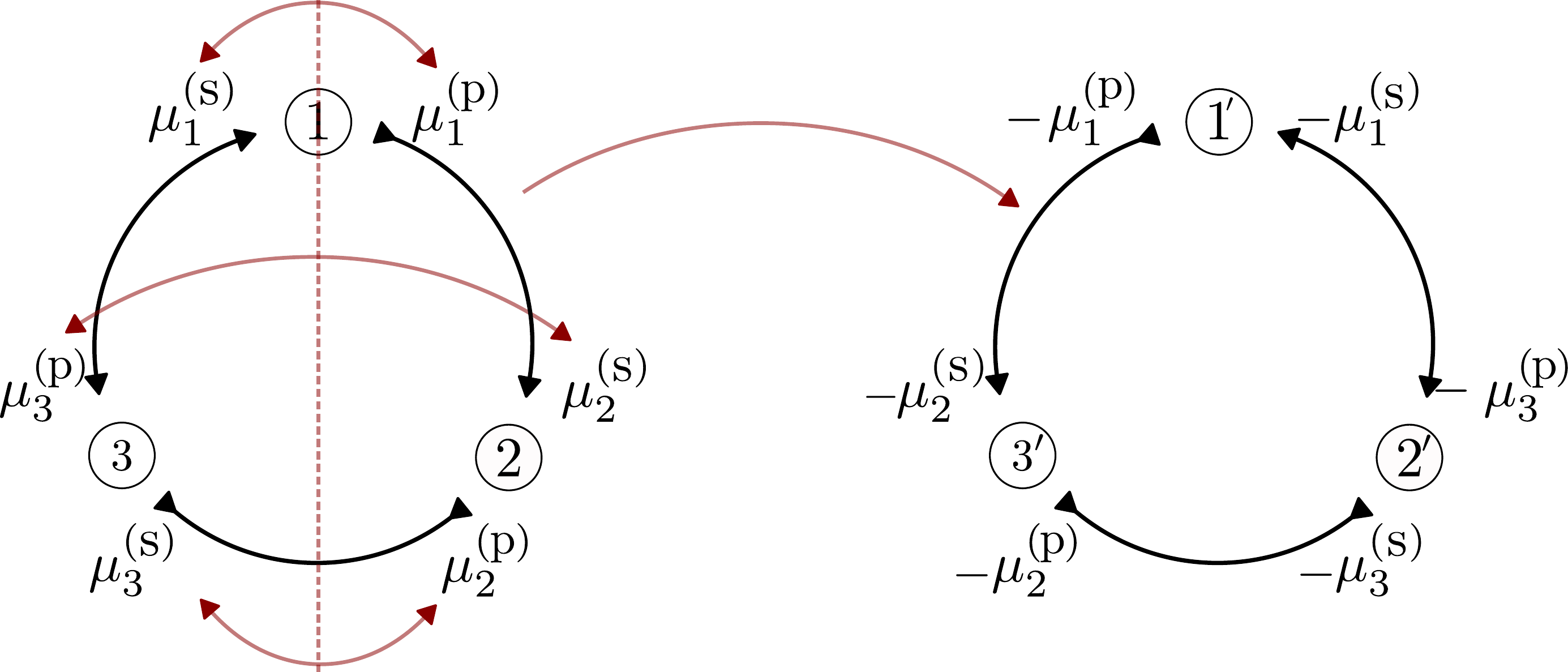}
\caption{Example of a reflection symmetry around pair $(2,3)$ 
applied to the network with pair interaction motifs AR--AA--RR,
yielding the network RR--AA--RA,
which is equivalent with respect to the second order instability condition.}
\label{sifig:mirrorsym}
\end{figure}

One key aspect making the classification of cyclic networks easier
is that the stability conditions are invariant under reflection symmetry around
a pair of species.
The corresponding symmetry operation is to swap the interaction network
around a reflection line.
For instance, applying a reflection symmetry around pair (2,3)
involves:
\begin{itemize}
    \item Turning $\mus_1$ into $-\mup_1$ and vice versa
    \item Turning $\mup_2$ into $-\mus_3$ and vice versa
    \item Turning $\mus_2$ into $-\mup_3$ and vice versa
\end{itemize}
with the sign change of the mobilities coming from the fact that substrate and product interactions of the same signs 
have opposite associated mobility signs.
See \cref{sifig:mirrorsym} for a graphical example.
Additionally, the activities and homogeneous densities of the species whose mobilities are swapped also need to be exchanged.

The stability of all possible networks of self-repelling species can then be determined by enumerating them and, for a
given network, applying the cyclic swap and reflection symmetry to determine all equivalent networks. We can group the
interaction networks of self-repelling species into three classes, based on whether they contain one, two, or three
unique motifs. We determine the stability of each of the retained elementary networks using the method described in
Section \ref{sec:pair_class},  and compile the results in \cref{tab:elem_nets}.

%

\end{document}

%% file: Tables/cycle_stability.tex
\newcommand{\Cm}{\textrm{C}_-}
\newcommand{\Cp}{\textrm{C}_+}
\newcommand{\au}[1]{{\color{NavyBlue}\textbf{#1}}}
\newcommand{\cu}[1]{{\color{Aquamarine}\textbf{#1}}}
\begin{table*}
    \centering
    \begin{tabular}{c@{\hspace{5em}}c@{\hspace{5em}}c@{\hspace{5em}}c}
                          \toprule
                Stability Class        &         \multicolumn{2}{c}{\hskip-2cm Interaction Motifs}        &                  Elementary Cycle \\
                                       &                         Self [11--22--33]                          &               Mutual [12--23--31]                     &                                            \\
                             \midrule
                Always UNSTABLE        &               \au{A} -- \au{A} -- \au{A}              &                  XX -- XX -- XX                 & \tablegraphics{0.1\textheight}{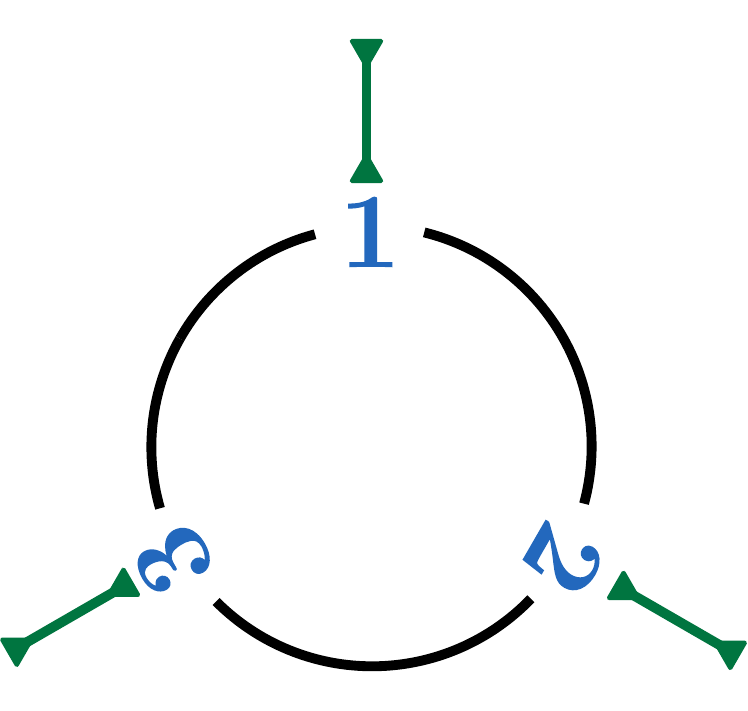}\\
                             \midrule
         Type-I UNSTABLE               &                    R -- \au{A} -- X                   &                  XX -- XX -- XX                 &       \tablegraphics{0.1\textheight}{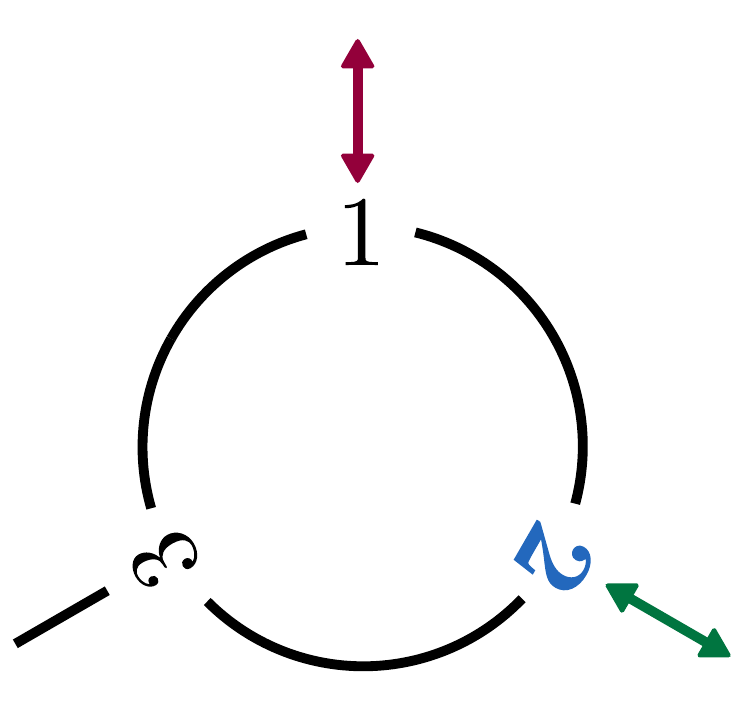}\\
                             \midrule
                                       &                       R -- R -- R                     &            AR  -- \au{AA} -- RR             &  \tablegraphics{0.1\textheight}{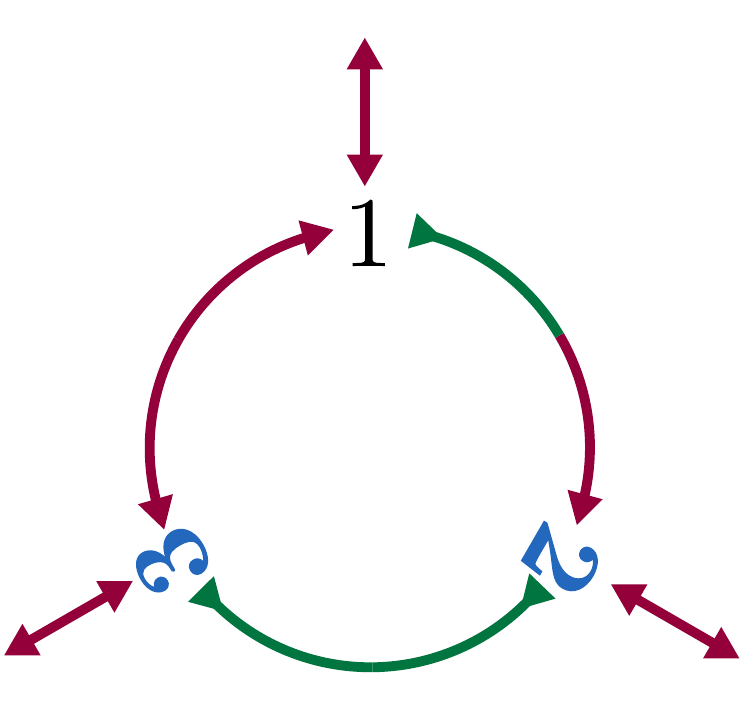} \\
         Type-IIa (strongly) UNSTABLE  &                       R -- R -- R                     &           AR -- \au{AA} -- RA           &  \tablegraphics{0.1\textheight}{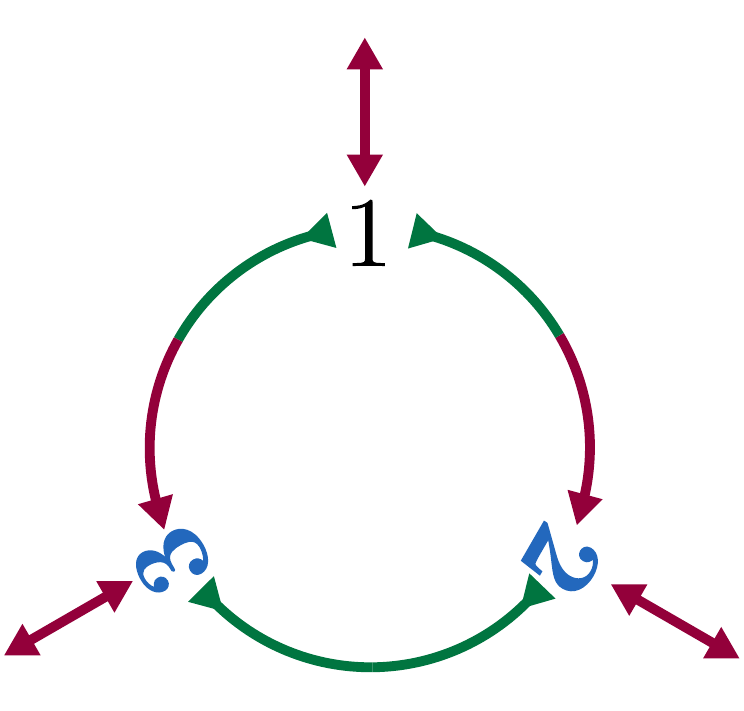} \\
                                       &                       R -- R -- R                     &         \cu{AA}  -- \cu{AA}  -- \cu{RR}         &  \tablegraphics{0.1\textheight}{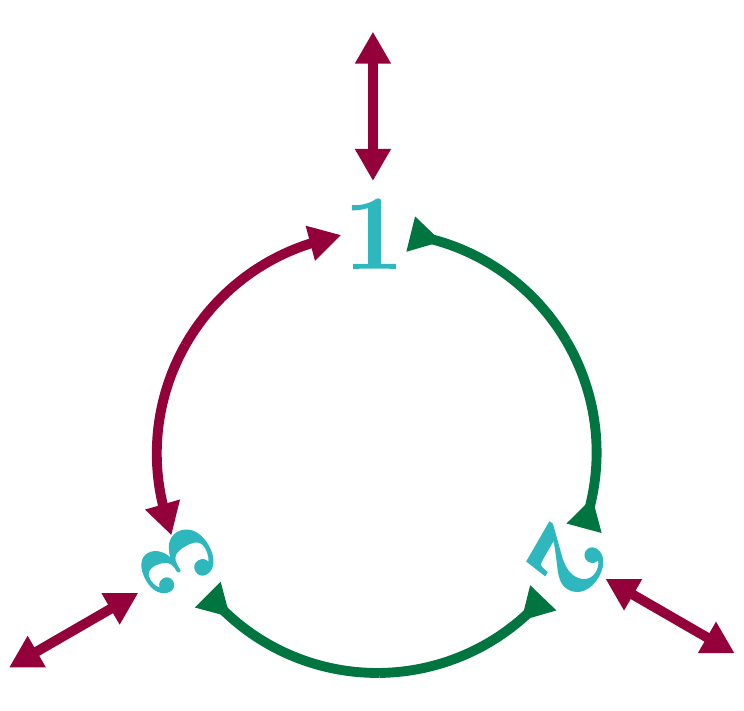} \\
                             \midrule
                                       &                       R -- R -- R                     &          \cu{AA}  -- RA -- RA           &  \tablegraphics{0.1\textheight}{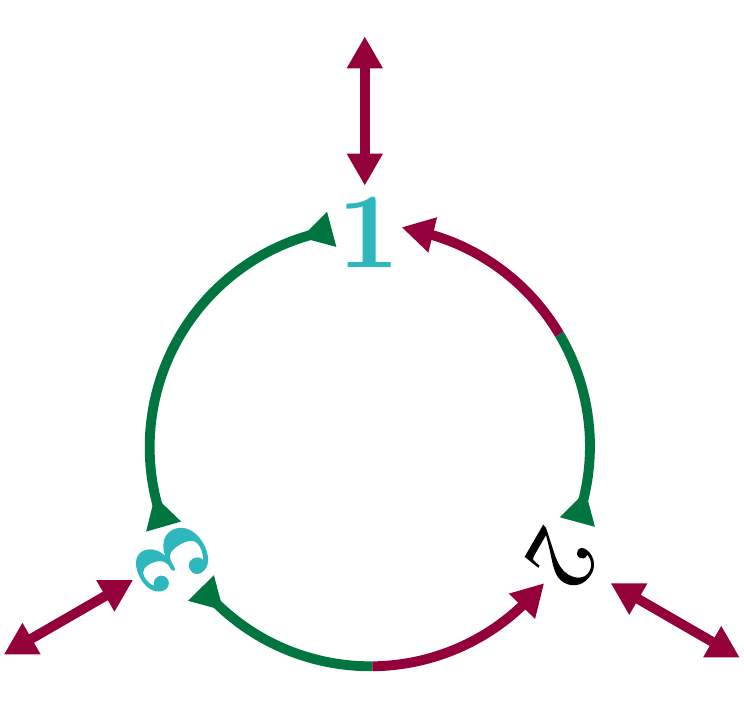} \\
         Type-IIb (weakly) UNSTABLE    &                                                       &                                              &                                                        \\
                                       &                       R -- R -- R                     &             RA -- AA -- \cu{AA}             &  \tablegraphics{0.1\textheight}{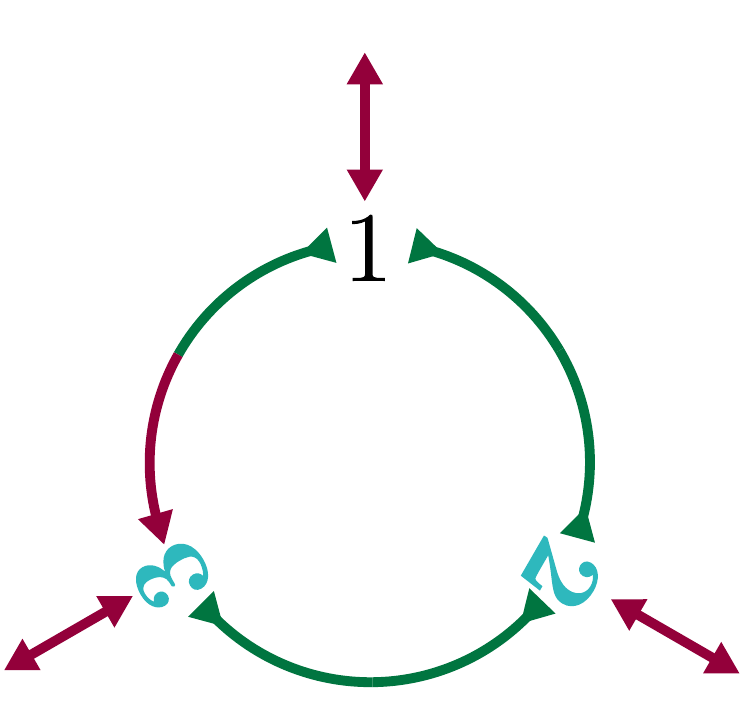} \\
                             \midrule
                                       &                       R -- R -- R                     &                 AA  -- AA -- AA                 &   \tablegraphics{0.1\textheight}{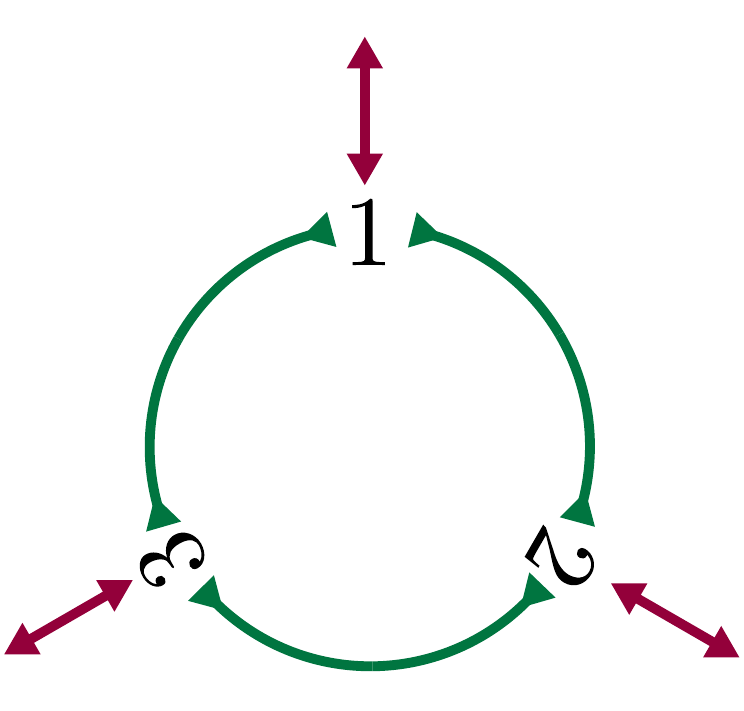} \\
                     STABLE            &                                                       &                                              &                                                        \\
                                       &                       R -- R -- R                     &            AR -- AR -- AR &  \tablegraphics{0.1\textheight}{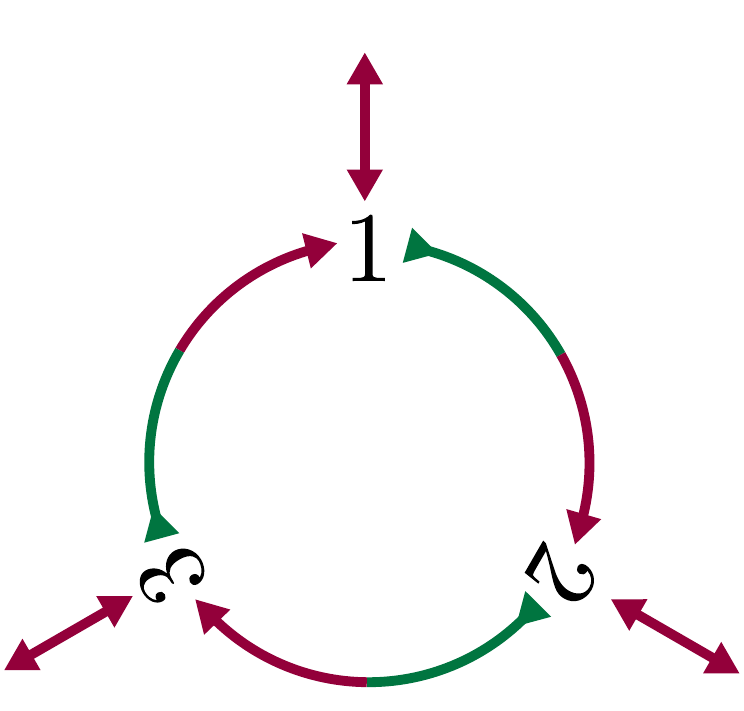}  \\
                       \bottomrule
    \end{tabular}
    \caption{
    classification of the different emergent possibilities for a size-3 metabolic cycle. Left column: Instability class; see main text for the definitions. Middle-left column: single-species interaction motifs, with A denoting self-attracting species and R denoting self-repelling species. X can correspond to any sense of interaction. Middle-right column: pair interaction motifs. Right column: corresponding metabolic cycle.
}
    \label{tab:elem_nets}
\end{table*}

%% file: Tables/pair_stability.tex
\begin{table*}[hbtp]
    \begin{tabular}{c@{ \hskip 0.5em }c@{ \hskip 1em }c}
\toprule
Motif                                                            &  \hspace{5.0em} Instability condition                                                 &  Stability contribution                 \\ \midrule
\tablegraphics{0.04\textheight}{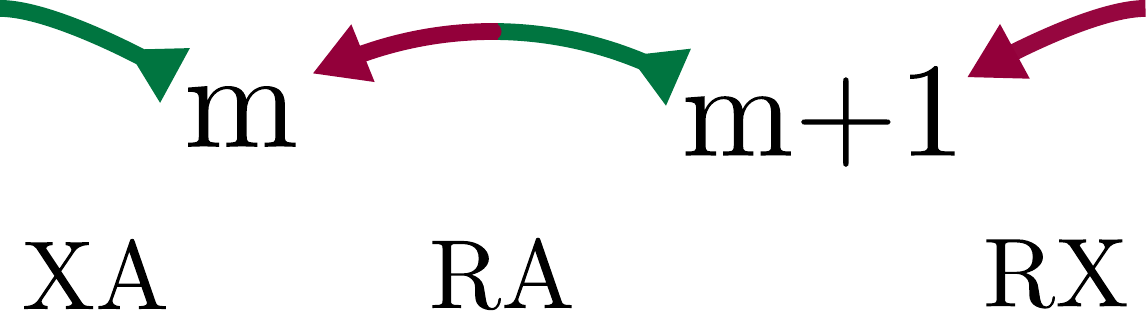} & \hspace{5.0em} $ \displaystyle \musovmup{m+1} + \mupovmus{m} < 1 $                   & Stabilizing \vspace{1.0em}\\
\tablegraphics{0.04\textheight}{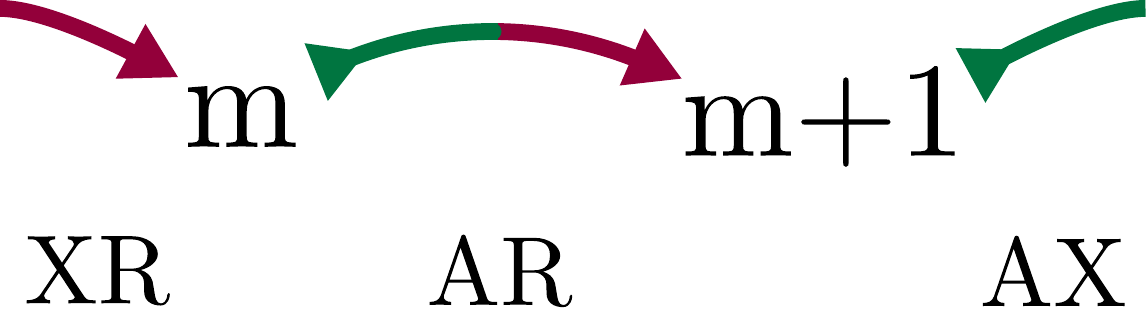}     &  \hspace{5.0em}$ \displaystyle \musovmup{m+1} + \mupovmus{m} < 1 $                   & Stabilizing \vspace{1.0em}\\
\tablegraphics{0.04\textheight}{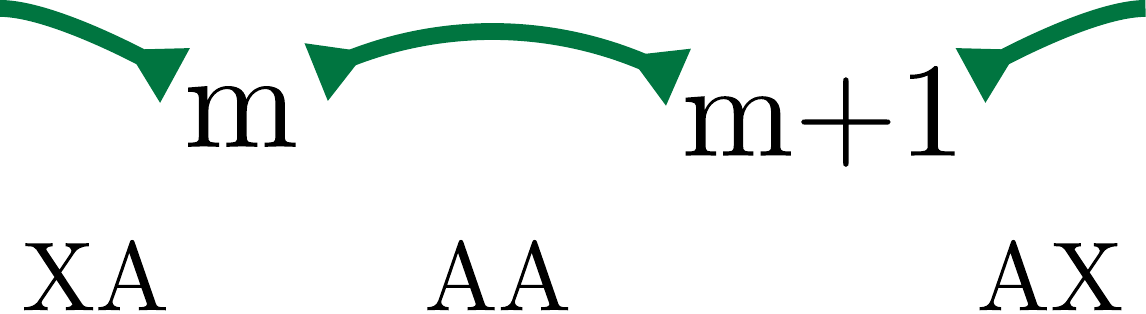}   & \hspace{5.0em} Trivially stable                                                      & Stabilizing \vspace{1.0em}\\
\tablegraphics{0.04\textheight}{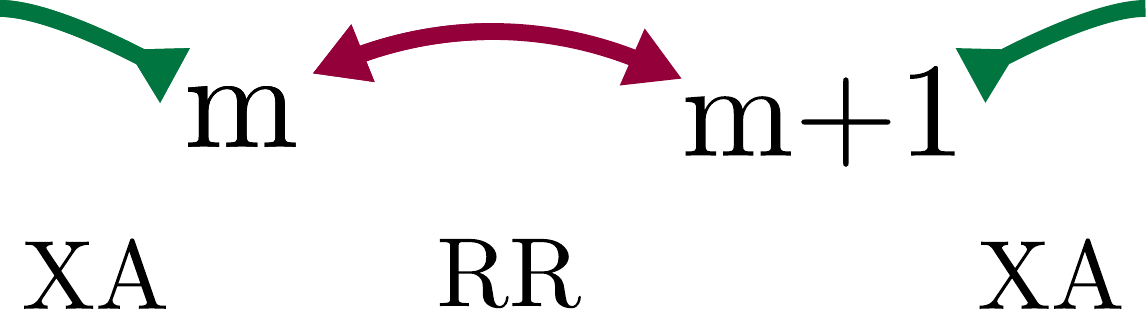}   &  \hspace{5.0em} $ \displaystyle \musovmup{m+1} + \mupovmus{m} >1 $                    & \hspace{5.0em}Conditionally destabilizing \vspace{1em}\\
\tablegraphics{0.04\textheight}{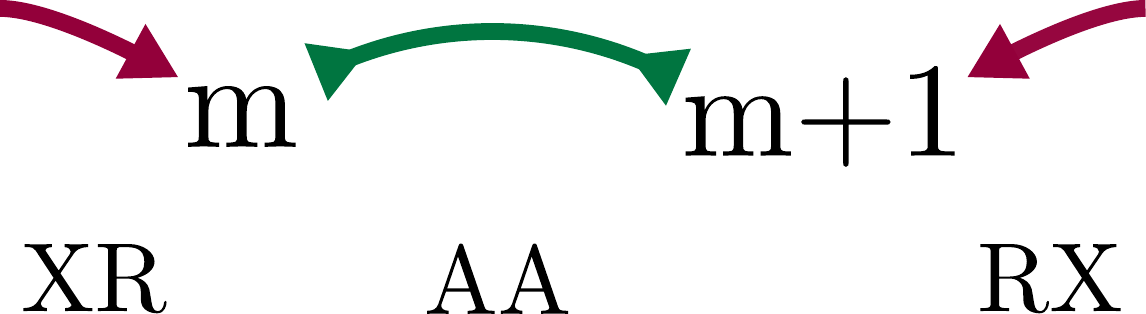}   & \hspace{5.0em} $ \displaystyle \musovmup{m+1} + \mupovmus{m} >1$                     & Destabilizing \vspace{1em}\\
\tablegraphics{0.04\textheight}{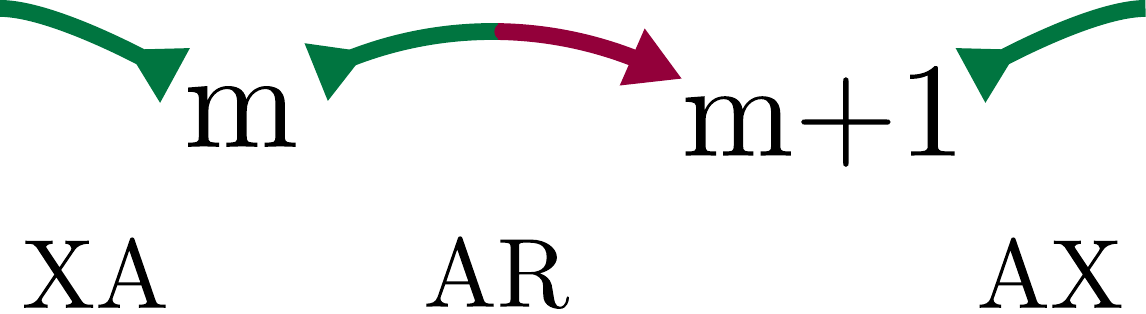}    & \hspace{5.0em} $ \displaystyle \mupovmus{m+1} \left( 1 - \mupovmus{m} \right) < 1 $  & Stabilizing \vspace{1em}\\
\tablegraphics{0.04\textheight}{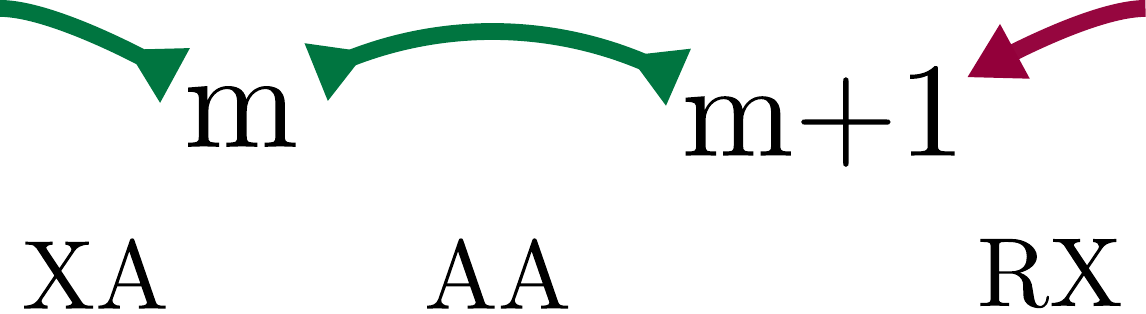}  & \hspace{5.0em} $ \displaystyle \mupovmus{m+1} \left( 1 - \mupovmus{m} \right) > 1 $  & \hspace{5.0em} Conditionally destabilizing \vspace{1em}\\
\tablegraphics{0.04\textheight}{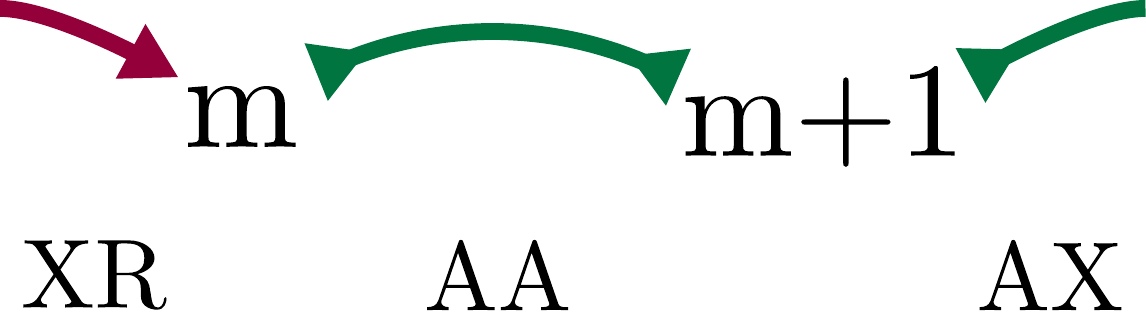}    & \hspace{5.0em} $ \displaystyle \musovmup{m} \left( 1 - \musovmup{m+1} \right) > 1 $  & \hspace{5.0em} Conditionally destabilizing \vspace{1em}\\
\tablegraphics{0.04\textheight}{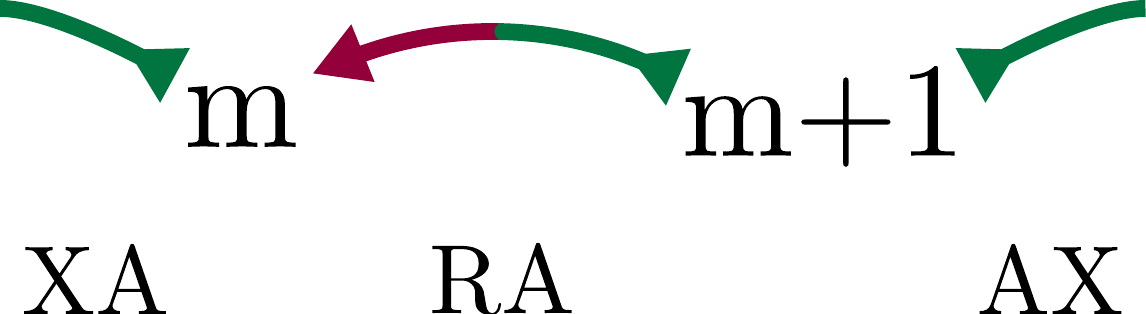}  & \hspace{5.0em} $ \displaystyle \musovmup{m} \left( 1 - \musovmup{m+1}\right)  < 1 $  & Stabilizing \vspace{1em}\\ \bottomrule
\end{tabular}
\caption{
    Stability of all self-repelling pair motifs.
    Left column: exhaustive enumeration of pair interaction motifs composed of self-repelling species 
    (labelled as ``Self R'' in \cref{fig:act_mob_cycle}(c)).
    ``A'' designates an attractive interaction, ``R'' corresponds to a repulsive interaction, and ``X'' can be any unspecified interaction.
    Middle column: condition on species mobilities for the corresponding term in \cref{eq:instab_2_terms} to be negative,
    i.e. instability-favouring.
    Right: contribution of the pair motif to the cycle stability, deduced from the corresponding middle-column condition (see Appendix \ref{sec:app_pair_stab} for details). A species pair is written as stabilizing if the associated condition cannot be satisfied, destabilizing if it 
     is always satisfied, and conditionally destabilizing if it the outcome depends on the mobility values.
}
\label{tab:motif_stab}
\end{table*}